# Recent Advances in Selective Image Encryption and its Indispensability due to COVID-19

Aditya Jyoti Paul [1,2],

1. Cognitive Applications Research Lab,
India.

2. Department of Computer Science and Engineering,
SRM Institute of Science and Technology,
Kattankulathur, Tamil Nadu – 603203, India.
aditya_jyoti@srmuniv.edu.in

*Abstract*— **The COVID-19 pandemic serves as a grim reminder of the unexpected nature of these outbreaks and gives rise to a unique set of research challenges in a variety of fields. As people all over the world adjust to this new 'normal', with most workplaces, from companies to educational institutions shifting online, enormous surges in the transmission of images and videos have been observed, creating record-breaking stresses on the internet backbone. At the same time, maintaining the privacy and security of the users' data is of immense importance, this is where fast and efficient image encryption algorithms play a vital role. This paper discusses the calamitous effects of the pandemic on the world population and how their changes in multimedia consumption have led to an urgent need for the development and deployment of secure and fast image encryption, especially selective image encryption techniques. It carefully surveys the most recent advances in this field, discusses their real-world effects and finally explores some future research avenues, to provide swift relief and recover from the disastrous effects of the pandemic.**

*Keywords— Chaos, COVID-19, Internet Usage Surge, Region of Interest, Multimedia encryption, Selective Image Encryption.*

## I. Introduction

At the time of writing this article, coronavirus has become one of the most severe maladies of recent times, claiming the lives of over 1 million people worldwide [1]. And these desperate times call for desperate measures, as people from all over the world come together to fight the pandemic, researchers from diverse fields like are faced with novel challenges born out of this scenario.

This respiratory disease, referred to as SARS-CoV-2 [2] or the Coronavirus Disease 2019 (COVID-19), originated in the Wuhan province of China, and was first identified in December 2019, it was declared a Public Health Emergency of International Concern on 30th January, 2020 and deemed to be a pandemic on 11th March, 2020 by WHO (World Health Organization), a detailed timeline of its responses can also be found in [3].

A systematic analysis of the evolution of COVID-19, analyzing data over a period of 100 days, was presented in [4]. Examining numerous detailed case-studies like those focusing on the worst-hit states in India [5], Heilongjiang province in China [6] and the world in general [7], various government policies were formulated. As a result of rising public health concerns, most governments established lockdowns to varying degrees of strictness, preventing people from leaving their homes or gathering in groups. This resulted in people feeling listless and bored, go through mental distress and much more. [8] and [9] discussed in detail some behavioral changes brought on by the spread of COVID-19, and its impact on health and well-being of individuals. A study [10] on 917 Chinese adults, assessed the state boredom and psychological distress among the people and assessed the multimedia consumption of these individuals.

Many quantitative studies like [11] and [12] were done to analyze the effects of the pandemic on factors including internet and screen usage, and [12] reports an almost two-thirds increase in usage of screen-time due to COVID19. In [13], the authors analyze the massive changes in markets like tourism, retail and higher education and the effects of the same on health, cash flow, marketing and social media and internet usage. [14] discussed a case-study of how the lockdown affected a medium-sized university's campus network, observing massive spikes in internet and multimedia usage due to virtual classrooms, live classes, remote online collaboration etc.

Such stresses on the internet backbone have already catalyzed improvements in the deployment of internet connectivity and governments can play an active role by implementing a variety of policy changes suggested in [15], like allocating the unassigned spectrum in a fair and non-discriminatory way, temporarily freezing of payment for internet and mobile data, supporting telecommunication infrastructure providers and monitoring network capacity regularly. However, such policy changes are subject to the societal and political contours of the specific region and in some cases somewhat difficult to implement due to legal constraints.

The impact of the surge in internet usage during the pandemic has been studied in great detail in [16], which also discusses some viewpoints for further research and practice. The article mentions some major research issues of which increasing digitalization and increasing work from home (WFH) practices are majorly responsible for the increase in multimedia consumption. For example, increased digitalization has given almost everyone access to basic internet usage and mobile phones and an increase in WFH implies an increase in online presentations and meetings, in which the communication is mostly through the exchange of multimedia.

Given the massive usage of social media and mobile multimedia [17] over the past decade, and especially with multimedia-focused social media services like Instagram [18] and Snapchat taking over older and more run-of-the-mill social networking sites like Facebook in growth [19], the importance of swift communication of multimedia has to be appreciated. COVID-19 has brought to the forefront, an existing research challenge and development of faster image encryption schemes that can be easily deployed in a variety of scenarios, from workstations to edge devices like handhelds is urgently needed. The algorithm should also give the deployment endpoint flexible control over the level of encryption based on scenario or content.

The rest of the article is structured as follows. Sec. II discusses some of the preliminaries of selective image encryption like what it means, block and stream ciphers, different algorithms and domains etc. Sec. III discusses the recent advances in selective image encryption, and finally Sec. IV gives the conclusion and discusses future research avenues.

## II. Preliminaries

### A. What is Selective ROI Image Encryption

It is common knowledge that different regions of an image do not encompass equal concentrations of important information. Image encryption researchers have strived to efficiently extract the Regions of Interest (ROI) in an image so that smarter or adaptive encryption can be done, paying more attention to ROI regions. This does not compromise on the security of the system but time can be saved as fewer iterations can be done on non-ROI parts of the input image.

For example, Fig. 1 shows the different stages of encryption of a 512x512 grayscale image of Lena (viridis colormap used for better visualization), as it passes through a number of steps for selective image encryption. First, the image has been divided into smaller blocks of 64x64, and the importance of each block is calculated by measuring the number of Edges. Then based on this importance, confusion is carried to varying degrees in these blocks, to produce the intermediary image, on which finally diffusion is carried out.

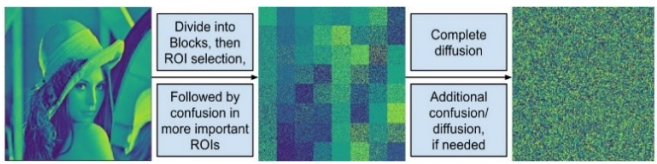

Fig. 1. An example flow of selective image encryption in practice.

ROI however is not the only way to carry out selective encryption, it can also be done by encrypting selected coefficients in the frequency domain, or hybrid domain, as has been discussed in Sec C and shown in Fig. 2, or even encrypting certain parts of the bit plane in all pixels, like only the *n* Most Significant bits (MSB) or the *n* Least Significant Bits (LSB).

### B. Block and Stream Ciphers

To decrypt an encrypted image E, the secret key may use either of the two sets of algorithms, block ciphers or stream ciphers. Block cipher is an example of symmetric key encryption, it is a deterministic and invertible function of *k*-bit keys and *n*-bit plaintext blocks to *n*-bit ciphertext blocks, choosing a suitable mode of operation like Electronic Code Book (ECB), Counter Mode (CTR), Cipher Block Chaining (CBC), Cipher Feedback (CFB) etc. each having different advantages and disadvantages.

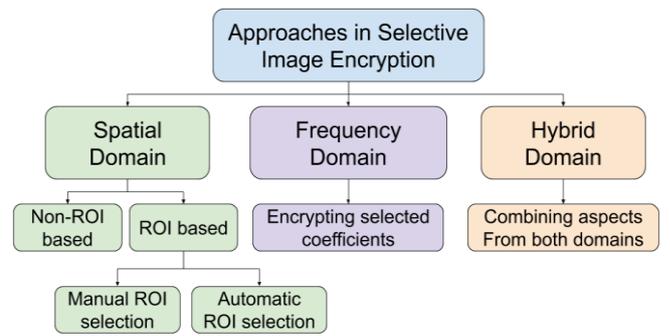

Fig. 2. Approaches of Selective Image Encryption in various domains.

A stream cipher is also an example of symmetric encryption, it maps *k*-bit keys and variable-length plaintexts to ciphertexts of the same length. Certain modes of block cipher actually produce synchronous stream ciphers like the CTR mode and the boundary between block and stream ciphers is often not rigid and has to be analyzed in careful detail paying attention to security, key integrity, runtime parallelization and more. A more detailed discussion on these ciphers can be found in [20].

### C. Encryption Algorithms and Domains

For symmetric key encryption, some standard algorithms have been extensively explored like Data Encryption Standard (DES), Triple DES (3DES), Advanced Encryption Standard (AES), Blowfish etc. While some popular stream cipher algorithms include RC4, SEAL, SNOW etc. The security of the encryption scheme also depends on the implementation as well.

Image encryption can be done in spatial domain (the image as is, with values in R, G, B channels and transparency), frequency domain (corresponding to the rate of change of pixel intensities) and also in hybrid domain, as shown in Fig. 2.

## III. Related Work

In the past few years, selective encryption has gained a lot of traction primarily due to its adaptability, flexibility for utilization in a variety of scenarios and good encryption effect in terms of security. Often when speed is key, it is not desirable to encrypt the entire image heavily. Selective encryption has been explored for close to two decades, but many challenges still remain like having an efficient key structure, giving the user control over the encryption level etc. In this section, the advances in the field of selective image encryption are carefully analyzed and compared in terms of speed, practicality, ease of deployment and suitability for use in various scenarios.

One of the earliest contributions to selective encryption was done by Xiang et al. [21] in 2007, wherein they proposed a scheme with spatiotemporal chaotic system from coupled map lattice and implemented it first for grayscale images and then extended it for encrypting RGB images as well. They also considered a selection strength parameter for the security time tradeoff as well, which was quite novel at that time. They reported their scheme to be very fast due to the leveraging of selective encryption and reported their scheme as being more efficient and universal compared to other block ciphers that are codec specific and related to image compression.

Another technique of selectively encrypting only the 4-bit MSB of each pixel was presented by Munir [22], where a logistic map was used to generate a chaotic keystream and a CBC-like mode was chosen for the algorithm. In [23], a selective bit-plane encryption technique based on cross-coupled chaotic tent map is proposed for both grayscale images and color images. Blowfish algorithm is used for selective image cryptography in [24], the authors report a better encryption performance compared to using DES and AES.

When discussing selective ROI based encryption, the process of choosing the ROIs can be manual or automatic (based on morphological features in the image). In [25], Panduranga et al. have shown both manual and automated applications of selective encryption on medical and satellite images, and in [26] partial image encryption is carried out using two-stage hill cipher technique. In [27], block-wise shuffling and 1D logistic map or tent map were utilized for partial encryption and security in terms of Net Pixel Change Rate (NPCR) and Unified Averaged Changed Intensity (UACI) seems to have massively improved. [28] also proposed a simple encryption scheme, involving manual selection, based on Logistic and Lorenz map for medical image encryption.

In [29], Bhatnagar and Wu proposed a simple and effective selective encryption technique based on pixels of interest and Singular Value Decomposition (SVD). In this method, the pixels are scrambled with the sawtooth space-filling curve (SFC), followed by diffusion on the significant pixels using a secret key obtained from a non-linear chaotic map and SVD. Choudhary et al. [30] proposed a hybrid approach involving splitting the image into blocks and confusion with Arnold cat map and repeated for varying block sizes. A selective encryption scheme was proposed by [31] which involved dividing an image into blocks as well and the TD-ERCS chaotic map.

Jawad et al. [32] in 2015 discussed a survey of the challenges in selective encryption of color images based on three categories, ROI selection, encryption techniques and key management, those works of literature have mostly not been repeated in this review to avoid duplication. Comprehensive surveys discussing the most recent advances in image encryption in general can also be found in [33] and [34].

In [35], a modification of T-DES encryption algorithm, combined with inverted LSB steganography was discussed for selective image encryption that was twice as fast as 3DES. In [36], a novel ROI selection based on saliency detection in compression domain was proposed, followed by encryption using parametric switching chaotic system and discrete wavelet transform based content transform (DWTCT). In [37], a selective encryption based on 2D DWT, Henon's chaotic map and 4D Qi hyperchaos is proposed. The 2D DWT is used to decompose the position of the pixels, which is shuffled with Henon's chaotic map, and then diffused by XORing with keystreams generated from the Qi attractor.

A study of the practical issues of Discrete Cosine Transform (DCT) based Selective Encryption for bitmap images was discussed in [38], where the effects of various parameters like protecting only a few coefficients in frequency domain vs protecting all the coefficients, the effect of rounding errors etc. They also proposed a method to guess DC coefficients from parts of AC coefficients, thus they note the importance of protecting the AC coefficients along with DC coefficients. Here ROIs were not used, but can be implemented as a pre-computation step for more effective encryption of those regions as well. Shakir proposed a selective AES coding of Haar wavelet transform with logistic chaos-based confusion in [39].

In [40], a Discrete Wavelet Transform (DWT) based selective image encryption scheme was implemented, that's compatible with the JPEG-2000 standard. It encrypted just 6.25% of the DWT coefficients and could resist medical inference from the images, even after an attack. A lightweight edge ROI-based selective medical image encryption scheme was proposed by Khashan et al. in [41]. Selective image encryption based on fractional wavelet domain was proposed by Taneja et al. in [42], encrypting only around 3.125% of the entire image data but providing acceptable levels of security.

In [43], a selective encryption scheme using square wave SFC [44] and Orthogonal Polynomials Transform had been proposed, which was incredibly fast and suitable for mobile usage. ROI-based techniques have also been demonstrated to be effective with SVD and chaos for copyright protection in [45]. In [46], selectively pixels are encrypted in spatial domain with Henon's chaotic map with key-value transformation. The paper also does a great comparison against some other ROI encryption schemes.

In recent times, [47] proposed selective medical image encryption with DNA cryptography for medical images and a combination of Chen's hyperchaotic system and Taylor Chirikov map. Similarly, DNA cryptography with wavelet transform was also used in [48] for selective encryption. [49] discussed the selective encryption of the DC and some AC coefficients using the number maze technique. The use of DNA cryptography for image encryption has also been demonstrated in [50-52]. Compressive sensing plays a vital role in fast image encryption, since the resulting images are smaller or equal in size as original image, it reduces network load and has been demonstrated in [53-56].

This section surveyed some of the most recent and notable works of literature in the field of selective image encryption which made use of various techniques like chaotic equations, space-filling curve for image iteration, DNA, numeric maze technique in various domains like spatial, frequency and hybrid domains and also used various key management techniques. The next section discusses some ongoing work in this field.

IV. OUR ONGOING RESEARCH

Having presented a survey of the recent advances, this section speaks in brief about some of the real-world issues that need to be solved, and showcases some of the ongoing research outputs of the author's experiments, to highlight the potential of selective image encryption and demonstrate how it will have a significant impact in easing the communication of secure multimedia, especially due to the surges caused by the COVID-19 Work from Home (WFH) scenario, across various industries.

The most essential research areas where this can be implemented right now will be in encrypting medical imagery, school or work multimedia and also other general images. The robustness and suitability of this field to tackle those problems, and some key ideas are discussed in the forthcoming sections.

### A. Encryption of Medical Images

The requirement in medical images encryption is not usually complete obfuscation, but making sure the attacker cannot obtain an image that can be used to get personally identifiable information (PII) or any diagnosable information about a patient. Here selective encryption will play an essential role in making sure the hospitals' internal and external communication is not overloaded due to long heavy encryption times of patient data as long all PII is removed.

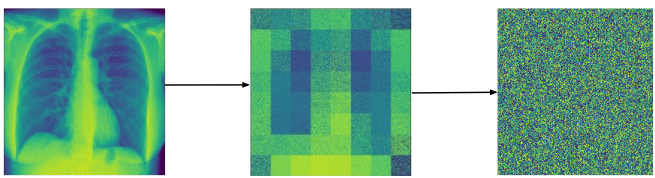

Fig. 3. ROI Block Cipher Image Encryption on chest radiograph.

Visual inspection of the steps of encryption of a chest radiograph in Fig. 3 shows ROI based image encryption is robust enough to protect against attacks on medical images, with even low security high speed settings. The images are in grayscale but viridis colormap has been used for better contrast and visualization of the encryption effect.

### B. Encryption of Educational and Office-work Multimedia

COVID-19 has forced most schools, colleges and higher degree institutes to shift online as the primary mode of pedagogy. It has also been shown that the use of multimedia helps achieve greater effectiveness in education [57]. Similarly, most workplaces, especially those with software roles have moved online, leading to a large amount of audio and multimedia information being transmitted over time.

Educational multimedia and office work multimedia will have different requirements of security and there will also be differences in terms of their software and hardware infrastructure. Thus, a company needing to transmit extremely sensitive multimedia, and having access to high-performance workstations might desire a higher level of security, which would certainly be different from the security level needed by people using their phones for learning or recreational activities.

These scenarios demand flexible algorithms that can be tuned to the needs of the organization or the deployment API, at a granular level, to tweak the security speed tradeoff parameter thus making sure their systems are as secure as needed and have maximum uptime.

In Fig. 4, the robustness of selective encryption is demonstrated on a screengrab from Google Meet, in which has the person's face and the slides are selectively encrypted, and the ROIs are automatically chosen by the system. Viridis colormap has been used on a grayscale image for improved contrast visualization and the screen size was adapted to 512x512 while creating the original input image to showcase the encryption effect on both the slides and the human face.

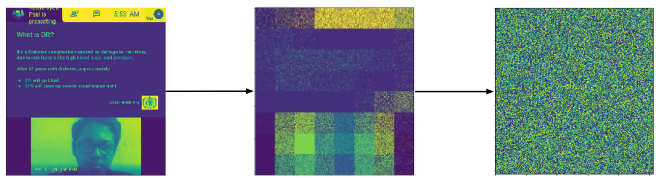

Fig. 4. ROI Block Cipher Image Encryption on Google Meet screengrab.

### C. Encryption of other general multimedia

Selective image encryption has the potential to be flexible enough for all multimedia and all deployment systems in general if implemented with consideration of the speed vs security parameter, and factors like efficient key structure, parallelization, etc. should also be carefully looked into. Some more ideas to make this generalizable are discussed in Sec V.

Organizations and even departments within the same organization, along with the users can have granular control over their data and security settings with this technology.

## V. FUTURE RESEARCH AVENUES

Having disseminated the current literature and shown the potential of selective image encryption this section discusses some of the potential avenues for future advancements in the field of selective image encryption and combining it with the advantages provided by advances in associated fields as well.

Due to COVID-19 and the rise of mobile multimedia in general, the goal right now is to have image encryption systems that do not clog networks hence it is of utmost importance to have the encrypted images of approximately the same size or lesser than the original images, compressive sensing has become a popular and efficient technique in that regard and has been utilized for image encryption in [58-60] in recent years.

One also needs to take care of the security aspects of the chaotic maps that are used in the encryption schemes. Small key spaces and poor initialization of chaotic maps can make the keystreams unchaotic. Numerous such analyses have been published in recent times like [61] which commented on the improper key space analysis done by [62] and experimentally identified the non-chaotic ranges. Similarly [63] proposed a novel metaheuristic approach based on Pareto evolution II to tune the hyperparameters of the chaotic maps and observed better security performance compared to traditional metrics.

Generalized image processing with domain adaptation is difficult [64] but with the rise of the Internet of Things (IoT), and federated learning [65-67], it is now possible to detect the image or video context on a device without any privacy risk, which can be utilized to adapt to the level of security needed, reducing response times. The image encryption system can also be sensitive to the software and hardware constraints and network loads in future to effectively balance security and speed. This would lead to having better and more efficient image encryption algorithms that also generalize well to various scenarios.

In ROI image encryption, it is of immense importance to make sure that ROIs that are being chosen are the most critical ones, the size of blocks in the partitioned image usually has an impact on security as well. These are some avenues selective image encryption researchers can pursue in the near future.

VI. CONCLUSION

This article discusses the enormous stresses placed on the internet backbone due to the WFH scenario and the coronavirus outbreak, causing massive surges in internet usage, which if left unchecked, can cause outages worldwide. Since multimedia consumption is at the crux of the problem, this survey goes over the most recent advances in the field of selective encryption, which is key to efficiently optimizing the speed and security tradeoffs and maintaining the required levels of privacy for user data. Selective encryption has been worked on for almost the past two decades, but still significant areas of improvement remain in areas like guaranteeing the security and integrity of the keystreams, giving the deployment endpoint or the user control over the level of encryption, or dynamically adapting to the software and hardware constraints etc. There is a need for standardized tests to compare algorithms across various architectures as well. Some more avenues of future research combining the advantages of federated learning and reduced response times from domain adaptation with selective encryption have already been explored in the previous section. The author hopes this survey provides a deeper insight into the image encryption landscape, especially in selective image encryption today, and encourages and enables researchers to solve some of the most pressing problems in this field.